\documentclass[a4paper]{jpconf}
\usepackage{accents}
\usepackage{iopams}
\usepackage{epsfig}
\usepackage{graphics}
\newcommand{\eeq}{\end{equation}}
\newcommand{\beq}{\begin{equation}}
\newcommand{\ba}{\begin{array}}
\newcommand{\ea}{\end{array}}
\newcommand{\bea}{\begin{eqnarray}}
\newcommand{\eea}{\end{eqnarray}}
\newcommand{\vev}[1]{\langle #1\rangle}

\newcommand{\eps}{\epsilon}

\newcommand{\al}{\alpha}

\newcommand{\mrm}[1]{\mbox{\rm #1}}
\newcommand{\rsb}[2]{\raisebox{#1}[0pt]{#2}}

\newcommand{\effn}[1]{\accentset{(#1)}}

\newcommand{\rd}{rd.\ }

\newcommand{\ra}{\rightarrow}

\begin{document}
\title{Running neutrino masses and mixing in a SU$(4)\times$SU$(2)^2
\times$U$(1)_X$ model}

\author{A Psallidas\footnote{Talk presented at the ``Corfu Summer
Institute'', Corfu-Greece,  September 4-14, 2005. Work in collaboration with T
Dent, G K Leontaris, and J Rizos.}}
\address{ Theoretical Physics Division, Ioannina University, GR-45110
Ioannina, Greece}

\begin{abstract} \noindent
 In this talk, we discuss the implications of the renormalization group equations
 for the neutrino masses and mixing angles in a supersymmetric string-inspired
 SU$(4)\times$SU$(2)_L\times $SU$(2)_R\times$U$(1)_X$ model with matter in fundamental
 and antisymmetric tensor representations only. The quark, charged lepton and neutrino Yukawa
 matrices are distinguished by different Clebsch-Gordan coefficients due to
 contracting over SU$(4)$ and SU$(2)_R$ indices. In order to permit for a more realistic,
 hierarchical light neutrino mass spectrum with bi-large mixing a second $U(1)_X$ breaking singlet with
 fractional charge is introduced. By numerical investigation we find a region
 in the model parameter space where the neutrino mass-squared
 differences and mixing angles at low energy are
 consistent with experimental data.
\end{abstract}

\section{Introduction}

The success of the minimal supersymmetric extension of the standard model theory
(MSSM) in explaining the low energy parameters is undisputed. The measured values
of the strong coupling constant $\alpha_{3},\alpha_{em}$ and the weak mixing
angle $\sin^2\theta_W$ are correctly predicted when the unification scale is
taken to be of the order $10^{16}$ GeV. Furthermore, neutrino oscillation
data~\cite{Bahcall:2004ut} suggest that the search for physics beyond the
successful Standard Model (SM) and MSSM is a natural step.

In particular, experimental facts imply intriguing relations between the large
mixing angles in the neutrino sector and the smaller ones in the quark sector.
For instance, the mixing angle $\theta_{\nu_{12}}$ and the Cabibbo mixing
$\theta_C$ could satisfy the so called Quark-Lepton Complementarity (QLC)
relation $\theta_{\nu_{12}}+\theta_C\approx \frac{\pi}4$~\cite{Petcov:1993rk}. It
has been shown~\cite{Minakata:2004xt} that this relation can be reproduced if
some symmetry exists among quarks and leptons, namely if in a unified framework
they form part of the same multiplet. Therefore, moving beyond the MSSM attempts
have been made~\cite{Antusch:2005ca} in order to realize QLC in the context of
models unifying quarks and leptons such as the Pati-Salam~\cite{Pati:1974yy}.
Apart from the QLC relations, a different possibility is offered from the
symmetry $L_e-L_{\mu}-L_{\tau}$, which implies an inverted neutrino mass
hierarchy and bimaximal mixing $\theta_{\nu_{12}}=\theta_{\nu_{23}}=\frac{\pi}4$,
with $\theta_{\nu_{13}}=0$~\cite{lelmlt}. This symmetry alone does not give a
consistent description of current experimental data, but
additional corrections and renormalization effects have still to be taken into
account. It has been shown~\cite{Leontaris:2004rd} in the context of MSSM
extended by a spontaneously broken $U(1)_X$ factor,
that the neutrino sector respects an $L_e-L_{\mu}-L_{\tau}$ symmetry. Small
corrections from other singlet vevs, which are usually present in a string
spectrum, can lead to a soft breaking of this symmetry and describe accurately
the experimental neutrino data.

Also, neutrino oscillation data imply that neutrino mass squared differences are
tiny, with $\Delta\,m^2_{sol} \approx 10^{-5}$ eV$^2$ and $\Delta\,m^2_{atm}
\approx 10^{-3}$ eV$^2$. The observed smallness of the neutrino masses as
compared to those of quarks and charged leptons finds an attractive explanation
in the seesaw mechanism. In order to realize this mechanism a theory beyond the
MSSM is needed that will incorporate the right-handed neutrinos and suppress
adequately the neutrino masses. Examples of higher symmetries including the SM
gauge group and incorporating the right-handed neutrino in the
 fermion spectrum, are the partially unified Pati--Salam model, based on
 $SU(4)\times SU(2)_L\times SU(2)_R$ , and the fully unified $SO(10)$.
 When embedded into perturbative string or D-brane models,
these may be extended by additional abelian or discrete fermion family
symmetries. Thus fermion masses and mixing angles can be compared to the
predictions of various types of models with full or partial gauge unification and
flavor symmetries.

An issue of utmost importance is the evolution of the neutrino parameters from
the high energy scale where the neutrino mass matrices are formed, down to their
low energy measured values, according to the renormalization group equations
(RGEs). Of course, one attempt could be to determine the neutrino mass matrices
from experimental data directly at the weak scale. Nevertheless, the Yukawa
couplings and other relevant parameters are not known at the unification scale. A
knowledge of these quantities at the unification mass could provide a clue for
the structure of the unified or partially unified theory and the exact (family)
symmetries determining the neutrino mass matrices at the GUT scale. In the
literature there have been made
attempts~\cite{Antusch:2005gp,Mei:2005qp,Ellis:2005dr} to determine the neutrino
mass parameters in a top-bottom or bottom-up approach.

 In this talk, we explore further~\cite{Dent:2006nm} the neutrino mass spectrum of a model with gauge symmetry
$SU(4)\times SU(2)\times SU(2)\times U(1)_X$ based on the 4-4-2
models~\cite{Antoniadis:1988cm,Allanach:1995ch,Allanach:1996hz},  whose
implications for quark and lepton masses were recently investigated
in~\cite{Dent:2004dn}. Various attractive features characterize these models.
Large Higgs representations, which are problematic to obtain in string models,
are not required. Also, third generation fermion Yukawa couplings are unified
\cite{AllanachKQuad} up to small corrections. Unification of gauge couplings is
allowed and, if one assumes the model embedded in supersymmetric string, might be
predicted \cite{LTracas}. Moreover, the doublet-triplet splitting problem is
absent.

 String derived models usually predict in the spectrum of the effective field
 theory model a large number of neutral singlet fields charged differently under
 the extra U$(1)_X$ symmetry. Some of them are required to obtain non-zero vevs of the order of the U$(1)_X$ breaking
 scale due to the D- and F-flatness conditions. In the present model,
 in order to describe accurately the low energy neutrino
 data we introduce two such singlets charged under
 U$(1)_X$~\cite{Irges:1998ax}. The previous model \cite{Dent:2004dn} with one
 such singlet could easily give rise to a spectrum of light neutrinos with
 normal hierarchy and bi-large mixing. However, after study of the
 renormalization group (RG) evolution and unification it was found that the
 scale of light neutrino masses too large to be compatible with observation.
 If we impose the correct scale of light neutrino masses, then some heavy
 right-handed neutrinos (RHN) would have masses above the unification scale,
 which is incompatible with our effective field theory approach.

 Therefore, from the superpotential three {\em a priori}\/
 independent expansion parameters appear, one coming from the Higgses which acquire v.e.v.'s at the
 SU$(4)\times$SU$(2)_R$ breaking scale and two from the singlets. Also, in the
 superpotential there may appear nonrenormalizable operators involving products
 of SU$(4)\times$SU$(2)_R$ breaking Higgses and, consequently, different contractions
 of gauge group indices are possible leading to different contributions
 depending on the Clebsch factors. We use a minimal set of nontrivial Clebsch
 factors to construct the Dirac mass matrices. Right-handed (SU$(2)_L$ singlet)
 neutrinos obtain Majorana masses through nonrenormalizable couplings to the
 U$(1)_X$-charged singlets and to Higgses, while light neutrinos will acquire
 masses via the see-saw mechanism.

 In this talk, we discuss the numerical solution of the renormalization group
 equations for the neutrino masses and mixing angles above, between and below the see-saw
 scales, for several sets of order 1 parameters which
 specify the heavy RHN matrix. In each case the results at low energy are
 consistent with current experimental data, and provide further predictions for
 the 1-3 neutrino mixing angle and for neutrinoless double beta decay.

\section{Description of the model}

In this section the noteworthy characteristics of the string inspired Pati-Salam
model extended by a U$(1)_X$ family-symmetry are been presented. The total gauge
group of the model is SU$(4)\times$SU$(2)_L\times$SU$(2)_R\times$U$(1)_X$. The
field content includes three copies of $(4,2,1)+(\bar 4,1,2)$ representations to
accommodate the three fermion generations $F_i+\bar F_i$ ($i=1,2,3$), \beq F_i =
\left(
\begin{array}{cccc} u_i& u_i&u_i&\nu_i
 \\
 d_i& d_i&d_i&e_i
\end{array} \right)_{\alpha_i},\
\bar F_i = \left( \begin{array}{cccc} u_i^c& u_i^c&u_i^c&\nu_i^c
 \\
 d_i^c& d_i^c&d_i&e_i^c
\end{array} \right)_{\bar\alpha_i} \label{fereps}\nonumber
\eeq
 where the subscripts $\alpha_i$, $\bar\alpha_i$ indicate the U$(1)_X$
 charge.
 So,  $F_i$ includes all left-handed SM fermions (quarks and leptons),
 whilst $\bar F_i$ contains their right-handed partners, including the
 right-handed neutrinos. Note that $F+\bar F$ makes up the {\bf16}-spinorial
 representation of SO$(10)$.
In order to break the Pati-Salam symmetry down to SM gauge group, Higgs fields
$H=(4,1,2)$ and $\bar H=(\bar 4,1,2)$ are introduced \bea H = \left(
\begin{array}{cccc} u_H& u_H&u_H&\nu_H
 \\
 d_H& d_H&d_H&e_H
\end{array} \right)_{x},
\; \bar H = \left( \begin{array}{cccc} u_{\bar H}^c& u_{\bar H}^c&u_{\bar
H}^c&\nu_{\bar H}^c

 \\
 d_{\bar H}^c& d_{\bar H}^c&d_{\bar H}&e_{\bar H}^c
\end{array} \right)_{\bar x}\label{Hireps}\nonumber
\eea which acquire vevs of the order $M_G$ along their neutral components \bea
\vev{H}= \vev{\tilde{\nu}_H} = M_{G},\
\vev{\bar{H}}=\vev{\tilde{\nu}_{\bar{H}}^c} = M_{G} \eea breaking the symmetry at
$M_{G}$:
\begin{equation}
\mbox{SU(4)}\times \mbox{SU(2)}_L \times \mbox{SU(2)}_R \longrightarrow
\mbox{SU(3)}_C \times \mbox{SU(2)}_L \times \mbox{U(1)}_Y. \label{422to321}
\end{equation}
The Higgs sector also includes the $h=(1,\bar{2},2)$ field which after the
breaking of the PS symmetry is decomposed to the two Higgs superfields of the
MSSM. Further, two $D=(6,1,1)$ scalar fields are introduced to give mass to color
triplet components of $H$ and $\bar{H}$ via the terms $HHD$ and
$\bar{H}\bar{H}D$~\cite{Antoniadis:1988cm}. The scalar fields $D$ and $h$ make up
the {\bf10} representation of SO$(10)$. Also, in the string version of the model
one encounters the following kinds of fractionally charged states:
$$ H'(4,1,1), \ \ \ \bar{H}'(\bar{4},2,1), \ \ \ h_L(1,2,1), \ \ \ h_R(1,1,2).   $$
 Such states usually create problems in the low energy effective theory. Due to
 the fact that the lightest fractionally charged particle is expected to be
 stable, if its mass is around the TeV scale, then the estimation of its relic abundances~\cite{ra} contradicts the upper
 experimental bounds by several orders of magnitude. However, the problem doesn't
 exist either if these states become massive at a high scale of the order of the
 heterotic string scale or if they become integrally charged modifying the
 hypercharge generator. Also, the problem can be solved constructing models
containing a hidden gauge group~\cite{Leontaris:1995sf} forcing the fractional
charged states to form bound states~\cite{eln}.

Under the symmetry breaking (\ref{422to321}) the following decompositions take
place :
 \beq
    {F}_L({\bf4},{\bf2},{\bf1})      \ra
     Q({\bf3},{\bf2},-\frac 16) + \ell{({\bf1},{\bf2},\frac 12)} \hspace{5.10cm}
     \nonumber$$$$
\bar{F}_R({\bf\bar 4},{\bf1},{\bf2}) \ra  u^c({\bf\bar 3}, {\bf1},\frac
23)+d^c({\bf\bar 3},{\bf1},-\frac 13)+
                           \nu^c({\bf1},{\bf1},0)+ e^c({\bf1},
                           {\bf1},-1)  \hspace{0.32cm}
                           \nonumber$$$$
\bar{H}({\bf\bar 4},{\bf1},{\bf2}) \ra  u^c_H({\bf\bar 3}, {\bf1},\frac
23)+d^c_H({\bf\bar 3},{\bf1},-\frac 13)+
                            \nu^c_H({\bf1},{\bf1},0)+
                             e^c_H({\bf1},{\bf1},-1) \hspace{-0.32cm}
                             \nonumber$$$$
{H}({\bf4},{\bf1},{\bf2}) \ra  u_H({\bf3},{\bf1},-\frac
23)+d_H({\bf3},{\bf1},\frac 13)+
              \nu_H({\bf1},{\bf1},0)+ e_H({\bf1},{\bf1},1)
              \nonumber$$$$
D({\bf6},{\bf1},1)   \ra  D_3({\bf3},{\bf1},\frac 13) + \bar{D}_3({\bf\bar
3},{\bf1},-\frac 13)
                             \hspace{4.5cm} \nonumber$$$$
h({\bf1},{\bf2},{\bf2})
  \ra  h^d({\bf1},{\bf2},\frac 12) + h^u({\bf1},{\bf2},-\frac 12)
 \hspace{4.5cm} \nonumber
\eeq where the fields on the left appear with their quantum numbers under the
Pati-Salam gauge symmetry, while the fields on the right are shown with their
quantum numbers under the SM symmetry.

Finally, two scalar singlet fields $\phi$, $\chi$ are introduced, charged under
U$(1)_X$ whose vevs will play an essential role in the fermion mass matrices
through non-renormalizable terms of the superpotential. In the stable SUSY vacuum
the two singlets obtain vevs to satisfy the D-flatness condition including the
anomalous Fayet-Iliopoulos term~\cite{DineSW}. The anomalous D-flatness
conditions allow solutions where the vevs of the conjugate fields $\bar\phi$ and
$\bar\chi$ are zero and we will restrict our analysis to this case. Note that in
general a string model may have more than two singlets and more than one set of
Higgses $H_i$, $\bar{H}_i$, with different U$(1)_X$ charge. All such fields may
in principle also obtain vevs,
however we find that two of them are sufficient to give a set of mass matrices in
accordance with all experimental data. Hence we consider any additional singlet
vevs to be significantly smaller.

A matter that should be dealt with caution is the breaking of the Pati-Salam
symmetry. The Higgses $H_i$, $\bar H_i$ may obtain masses through $H\bar{H}\phi$,
$H\bar{H}\chi$ and $H\bar{H}\phi\chi$ couplings. However, in order to break the
Pati-Salam group while preserving SUSY we require that one $H$-$\bar{H}$ pair be
massless at this level. This ``symmetry-breaking'' Higgs pair could be a linear
combination of fields with different U$(1)_X$ charges, which would in general
complicate the expressions for fermion masses.
The chiral spectrum is
summarized
in Table~\ref{spec}. We choose the charge of the Higgs field $h$ to be
$-\alpha_3-\bar\alpha_3$ so that that the 3\rd generation coupling
$F_3\bar{F}_3h$ is allowed at tree-level.

\begin{table}[here]
 \caption{Field content and U$(1)_X$ charges}\label{table1}
\begin{center}
\begin{tabular}{ccccc}
\br
         &  SU$(4)$ & SU$(2)_L$ & SU$(2)_R$ & U$(1)_X$  \\
         \mr
$F_i$      &     4    &     2     &     1     &  $\al_i $ \\
$\bar{F}_i$& $\bar{4}$&     1     & $\bar{2}$ & $\bar{\al}_i$ \\
$H$        &     4    &     1     &     2     &   $x$     \\
$\bar{H}$  & $\bar{4}$&     1     & $\bar{2}$ & $\bar{x}$ \\
$\phi$     &     1    &     1     &     1     &   $z$     \\
$\chi$     &     1    &     1     &     1
  &   $z'$     \\
$h$        &     1    & $\bar{2}$ &     2     & $-\al_3-\bar{\al}_3$ \\
$D_1$      &     6    &     1     &     1     &  $-2x$    \\
$D_2$      &     6    &     1     &     1     & $-2\bar{x}$ \\ \br
\end{tabular}
\end{center}\label{spec}
\end{table}

Now we proceed to terms in the superpotential which can give rise to fermion
masses. Charged fermions obtain only Dirac type mass terms, whilst neutral ones
may obtain Dirac and Majorana masses. Dirac type mass terms arise after
electroweak symmetry-breaking from couplings of the form
 \beq
 {\cal W}_{D} = y_0^{33} F_3\bar{F}_3h + F_i\bar{F}_j h
 \left( \sum_{m>0} y_m^{ij} \left(\frac{\phi}{M_U}\right)^m+ \right. \\
\left.\sum_{m'>0} y_{m'}^{ij} \left(\frac{\chi}{M_U}\right)^{m'}+ \sum_{n>0}
{y'}_n^{ij}  \left(\frac{H\bar{H}}{M_U^2} \right)^n+   \right.$$$$ \left.
\sum_{k,\ell>0} y_{k,\ell}^{ij} \left(\frac{\phi}{M_U}\right)^k
\left(\frac{\chi}{M_U}\right)^{\ell} +  \sum_{p,q>0} y_{p,q}^{ij}
\left(\frac{H\bar{H}}{M_U^2} \right)^p
\left(\frac{\phi}{M_U}\right)^{q}   \right.\\
 \left.+  \sum_{r,s>0} y_{r,s}^{ij}
\left(\frac{H\bar{H}}{M_U^2}
 \right)^r
\left(\frac{\chi}{M_U}\right)^{s}
  + \cdots \right) \label{Dsup}
\eeq
where $m$, $m'$, $n,k,l,p,q,r,s$ are appropriate integers. Apart from the
heaviest generation, all masses arise at non-renormalizable level, suppressed by
powers of the fundamental scale or unification scale $M_U$. The couplings
$y_{m,m'}^{ij}$, ${y'}_n^{ij}$ {\em etc.}\ are non-vanishing and generically of
order $1$ whenever the U$(1)_X$ charge of the corresponding operator vanishes,
thus:
 \beq
 \alpha_i-\alpha_3 +\bar{\alpha}_j-\bar{\alpha}_3 =  \{ -m z, -m' z' , -n (x+\bar{x}), \\
 -k z - \ell z' , -p (x+\bar{x})  -q z , -r (x+\bar{x}) -s z' \}. \nonumber
 \eeq
Other higher-dimension operators may arise by multiplying any term by factors
such as $(H\bar{H})^\ell\phi^s/M_U^{2\ell+s}$
where $\ell(x+\bar{x})+s\,z=0$. Such terms are negligible unless the leading term
vanishes.

Neutrinos
may in addition receive also Majorana type masses. These arise from the operators
\beq
 W_{M} = \frac{\bar{F}_i\bar{F}_j HH}{M_U} \left( \mu_0^{ij} + \sum_{t>0}
\mu_t^{ij} \left(\frac{\phi}{M_U}\right)^t  \right. \\
\left.+ \sum_{t'>0} \mu_{t'}^{ij} \left(\frac{\chi}{M_U}\right)^{t'} + \right.
$$$$ \left.\sum_{w>0} {\mu'}_w^{ij} \left(\frac{H\bar{H}}{M_U^2}\right)^w \right.
+
\\ \left. \sum_{k',\ell'>0} \mu_{k',\ell'}^{ij}
\left(\frac{\phi}{M_U}\right)^{k'}
\left(\frac{\chi}{M_U}\right)^{\ell'}  \right. \\
\left.+
 \sum_{p',q'>0} \mu_{p',q'}^{ij}  \left(\frac{H\bar{H}}{M_U^2}
 \right)^{p'}
\left(\frac{\phi}{M_U}\right)^{q'} \right. + $$$$ \left.  \sum_{r,s>0}
\mu_{r',s'}^{ij} \left(\frac{H\bar{H}}{M_U^2}
 \right)^{r'}
\left(\frac{\chi}{M_U}\right)^{s'}
  + \cdots
\right).
 \label{MM}
\eeq
 where $t$, $t'$ , $w,k',\ell',p',r',s'$ are appropriate integers.
 Couplings of this type are non-vanishing whenever the following
 conditions are satisfied:
\beq
 \bar{\alpha}_i+\bar{\alpha}_j +2x = \{ - tz, -t' z', -w (x+\bar{x}), \\
-k' z - \ell' z' -p' (x+\bar{x}) -q' z, -r' (x+\bar{x}) -s' z' \}. \nonumber
\eeq

\section{Fermion mass matrices}

\subsection{General structure}
\

As can be seen from the superpotential Yukawa couplings (\ref{Dsup}) and
(\ref{MM}), three different expansion parameters appear in the construction of
the fermion mass matrices. These are \bea \eps\equiv \frac{\vev{\phi}}{M_U},\ \ \
\eps'\equiv \frac{M_{\rm G}^2}{M_U^2},\ \ \ \eps''\equiv \frac{\vev{\chi}}{M_U}
\label{expanspar} \eea given $\vev{H\bar H}= M _{G}^2$. Note that, for
non-renormalizable Dirac mass terms involving several products of
$H\bar{H}/M_U^2$, the gauge group indices may be contracted in different ways
\cite{Allanach:1996hz}.  This can lead to different contributions to the up, down
quarks and charged leptons, depending on the Clebsch factors
$C^{ij}_{n(u,d,e,\nu)}$ multiplying the effective Yukawa couplings.  Also,
although the Clebsch coefficient for a particular operator $O_n$ may vanish at
order $n$, the coefficient for the operator $O_{(n+p);q}$ containing $p$
additional factors $(H\bar{H})$ and $q$ factors of $\phi$ and/or $\chi$ is
generically nonzero.

In our analysis we wish to estimate the effects of the second singlet ($\chi$)
contributions on the neutrino sector as compared to the analysis presented
in~\cite{Dent:2004dn} without affecting essentially the results in the quark
sector. In order to obtain a set of fermion mass matrices with the minimum number
of new operators, we assume fractional $U(1)_X$ charges for $H,\bar H$ and $\chi$
fields, while the combination $H\bar H$ and the singlet $\phi$ are assumed to
have integer charges.
Thus $\alpha_i$, $\bar{\alpha}_i$, $x+\bar x$ and $z$ are integers, while $z'$,
$x$ and $\bar{x}$ are fractional. As a result, the Dirac mass terms involving
vevs of $\chi$ are expected to be subleading compared to other terms. Suppressing
higher-order terms involving products of $\eps,\eps'$ and $\eps''$, the Dirac
mass terms at the unification scale are
 \beq
 m_{ij} \approx \delta_{i3}\delta_{j3}m_3  + \left( \eps^{m} +
(\eps'')^{m'}+ C_{ij} (\eps')^{n} \right)\,v_{u,d}
 \eeq
where $m_3\equiv v_{u,d} y_0^{33}$, with $v_u$ and $v_d$ being the up-type and
down-type Higgs vevs respectively, and we omit the order-one Yukawa coefficients
$y^{ij}_m$ {\em etc.}\/ for simplicity.


 The Majorana mass terms are proportional to the combination $HH$
 (see Eq.~(\ref{MM})) which has fractional $U(1)_X$ charge. Thus,
 terms proportional to $\chi/M_U$ become now important for the structure
 of the mass matrix. The general form of the Majorana mass matrix is then
  \beq
 M_N \approx M_R \left( \mu_t^{ij}\epsilon^{t} +
 \mu_{t'}^{ij}(\epsilon'')^{t'} + {\mu'}_{w}^{ij} (\epsilon')^{w}  \right. \nonumber   \\
\left.  + \mu_{k',l'}^{ij} \epsilon^{k'} (\epsilon'')^{l'} + \mu_{p',q'}^{ij}
(\epsilon')^{p'} \epsilon^{q'} + \mu_{r',s'}^{ij} (\epsilon')^{r'}
(\epsilon'')^{s'}
 \right)
 \eeq
where we define $M_R\equiv M_{\rm G}^2/M_U\equiv\eps'M_{U}$.


\subsection{Specific choice of $U(1)_X$ charges}
\

Before we proceed to a specific, viable set of mass matrices, we first make use
of the observation~\cite{Dent:2004dn} that the form of the fermion mass terms
above is invariant under the shifts
 \beq
\alpha_i \rightarrow \alpha_i + \zeta, \ \bar{\alpha}_i \rightarrow
\bar{\alpha}_i + \bar{\zeta}, \ x \rightarrow x - \bar{\zeta}, \ \bar{x}
\rightarrow \bar{x} + \bar{\zeta} \label{zeta}
 \eeq
so that we are free to assign $\alpha_3=\bar\alpha_3=0$. We further fix
$x+\bar{x}=1$ and $z=-1$. The resulting $U(1)_X$ charges are presented in Table
\ref{charges}.
\begin{table}[!h]
\caption{\label{charges} Specific choice of U(1)$_\mathrm{X}$ charges.}
 \begin{center}
 \begin{tabular}{cccccccccccc}
    \br
  field & $F_1$ & $F_2$ & $F_3$ & $\bar{F}_1$ & $\bar{F}_2$ &
$\bar{F}_3$ &
$h$ & $H$ & $\bar H$ & $\phi$ & $\chi$  \\
    \mr
U$(1)_X$& -4 &  -3 &  0 &
 -2 &  1 & 0 &
 0 & $x$ &  $\bar{x}$ & -1 & $z'$ \\
    \br
\end{tabular}
\end{center}
\end{table}
We will choose the values of $x$ and $z'$ to be fractional such
that the v.e.v.\ of $\chi$ only affects the overall scale of neutrino masses, as
explained below.

The charge entries of the common Dirac mass matrix for quarks, charged leptons
and neutrinos
are then \beq \label{chargem} Q_X[M_D] =
\left( \begin{array}{ccc}
-6 & -3 & -4 \\ -5 & -2 & -3 \\ -2 & 1 & 0
\end{array} \right),
\eeq and the charge matrix for heavy neutrino Majorana masses is \beq Q_X[M_N] =
2x +
\left( \begin{array}{ccc}
-4 & -1 & -2 \\ -1 & 2 & 1 \\ -2 & 1 & 0
\end{array} \right)
\label{pmatrix}
\eeq Now, we relate $\eps,\eps',\eps''$ with a single expansion parameter $\eta$,
assuming the relations
 \beq
\eps =b_1 \sqrt{\eta},\ \eps''\equiv b_2 \eta,\ \eps'=\sqrt{\eta} \label{rel}
 \eeq
where $b_1$, $b_2$ are numerical coefficients of order one. Then the effective
Yukawa couplings for quarks and leptons may include terms \beq Y_f^{ij}=
 b_1^m \eta^{m/2} + b_1^{m+1} \eta^{1+m/2}
+ C_f^{ij}\eta^{n/2} + b_1 \eta^{1+n/2} + \cdots
 \label{yukawas}
\eeq with $f=u,d,e,\nu$, up to order 1 coefficients $y_f^{ij}$. Which of these
terms survives, depends on the sign of the charge of the corresponding operator.
For a negative charge entry, the first two terms are not allowed and only the
third and fourth contribute. Further, if a particular $C_{f}^{ij}$ coefficient is
zero, then we consider only the fourth term.

Therefore, we need to specify the Clebsch-Gordan coefficients $C^{ij}_f$ for the
terms involving powers of
$\langle{H\bar H}\rangle/M_U^2$. These coefficients
could be found if the fundamental theory
was completely specified at the unification or string scale. In the absence of a
specific string model, here we present a minimal number of operators which lead
to a simple and viable set of mass matrices. Up to possible complex phases, we
choose $C_d^{12}=C_d^{22}=\frac 13$, $C_u^{23}=3$ and
$C_u^{11}=C_u^{12}=C_u^{21}=C_u^{22}=C_u^{31}=C_{\nu}^{22}=C_{\nu}^{31}=0$ with
all others being equal to unity. The effective Yukawa matrices at the GUT scale
obtained under the above assumptions are
 \beq
 Y_u = \left( \begin{array}{ccc}
 b_1 \eta^{4} & b_1 \eta^{5/2} & \eta^2 \\
  b_1 \eta^{7/2} & b_1 \eta^{2} & 3\eta^{3/2} \\
 b_1 \eta^{2} & b_1 \eta^{1/2} & 1
\end{array} \right),\
 Y_d = \left( \begin{array}{ccc}
 \eta^3 & \frac{\eta^{3/2}}{3} & \eta^2 \\
 \eta^{5/2} & \frac{\eta}{3} & \eta^{3/2} \\
 \eta & b_1  \eta^{1/2} & 1
\end{array} \right), \nonumber $$$$
 Y_e = \left( \begin{array}{ccc}
 \eta^3 & \eta^{3/2} & \eta^2 \\
 \eta^{5/2} & \eta & \eta^{3/2}  \\
 \eta & b_1  \eta^{1/2} & 1
\end{array} \right),\
 Y_{\nu} = \left( \begin{array}{ccc}
 \eta^3 & \eta^{3/2} & \eta^2 \\
 \eta^{5/2} & b_1 \eta^{2} & \eta^{3/2}  \\
 b_1 \eta^{2} & b_1 \eta^{1/2} & 1
\end{array} \right).
\eeq
 where we suppress order one coefficients.  The quark sector as well as the
neutrino sector were studied
 in~\cite{Dent:2004dn}. However, full renormalization group
 effects
 were not calculated for the neutrino sector and as it turns out
 one singlet is inadequate to accommodate the low energy data.
 Consequently, we introduced the second singlet $\chi$, with
 fractional charge, whose v.e.v.\ affects only the overall scale
 of neutrino masses.

The desired matrix for the right handed
 Majorana neutrinos may result from more than one choice of
 charge for the $H$ field and the $\chi$ singlet field. These can
 be seen in Table \ref{posscharges}.

\begin{table}[!h]
\begin{center}
\caption{\label{posscharges} Possible choices for the U(1)$_\mathrm{X}$ charges
of the $H$ and $\chi$ fields, namely $x$ and $z'$ respectively.}
\begin{tabular}{cccccccc}
    \br
  $Q_X[H]$ & $-\frac 23$ & $-\frac 43$ & $-\frac 53$ & $-\frac 65$ & $-\frac 78$ &
$-\frac 14$ &
$-\frac 34$   \\
    \mr
$Q_X[\chi]$& $-\frac 53$ &  $-\frac 13$ & $\frac 13$ &
 $-\frac 35$ &  $-\frac 54$ & $-\frac 52$ &
 $-\frac 32$  \\
    \br
\end{tabular}
\end{center}
\end{table}

 We choose the $H$ charge to be $x=-\frac 65$ so that $2\,x$
 is non-integer, and set the
 $\chi$ singlet charge to $-\frac 35$. The analysis for the
 quarks and charged leptons remains the as in~\cite{Dent:2004dn}
 since operators with nonzero powers $\chi^r$
 do not exist for powers $r<5$ and are
 negligible compared to the leading terms. 

 With these assignments, the charge entries of the
 heavy Majorana matrix Eq.~(\ref{pmatrix}) are:
 \beq
 Q_X[M_N] =\left( \begin{array}{ccc}
 -\frac{32}{5}  & -\frac{17}{5} & -\frac{22}{5} \\
  -\frac{17}{5} & -\frac{2}{5}& -\frac{7}{5} \\
   -\frac{22}{5} & -\frac{7}{5}  & -\frac{12}{5}
\end{array} \right).
 \eeq
 Due to the fractional $U(1)_X$ charges
 contributions from $\phi$ or $H\bar{H}$
 alone vanish. However, we also have the singlets $H\bar{H}\chi/M_U^3$ with
 vev $b_2\, \eta^{3/2}$ and $\phi \chi /M_U^2$ with a vev $b_1 b_2 \,\eta^{2}$,
 while for some entries one may have to consider higher order
 terms since the leading order will be vanishing.
 In Table \ref{operators} we explicitly write the operator for
every entry of $M_N$.
\begin{table}[!t]
\caption{\label{operators}Operators producing the Majorana right handed neutrino
matrix $M_N$.}
\begin{center}
\begin{tabular}{ccc}\br
 $M_N$ entry & Operator & vev\\ \mr
 $  M_N^{11}$ & $ \left(\frac{H\bar{H}}{M_U^2}\right)^7 \frac{\chi}{M_U} $
 & $b_2 \eta^{9/2}$\\
 $  M_N^{12}$ & $  \left(\frac{H\bar{H}}{M_U^2}\right)^4 \frac{\chi}{M_U} $
 & $b_2 \eta^{3}$\\
 $  M_N^{13}$ & $ \left(\frac{H\bar{H}}{M_U^2}\right)^5 \frac{\chi}{M_U} $
 & $b_2 \eta^{7/2}$ \\
 $  M_N^{22}$ & $ \left(\frac{H\bar{H}}{M_U^2}\right) \frac{\chi}{M_U} $
 & $b_2  \eta^{3/2}$\\
 $  M_N^{23}$ & $ \left(\frac{H\bar{H}}{M_U^2}\right)^2 \frac{\chi}{M_U} $
 & $b_2 \eta^{2} $ \\
 $  M_N^{33}$ & $ \left(\frac{H\bar{H}}{M_U^2}\right)^3 \frac{\chi}{M_U} $
 & $b_2 \eta^{5/2}$\\
\br
\end{tabular}
\end{center}
\end{table}
  The Majorana right-handed neutrino
 mass matrix is then
 \beq \label{MRHN}
 M_N=\left( \begin{array}{ccc}
 \mu_{11} \eta^{9/2}  & \mu_{12}\eta^{3} & \mu_{13}\eta^{7/2}\\
 \mu_{12}\eta^{3} & \mu_{22}\eta^{3/2}& \mu_{23}
 \eta^{2} \\
  \mu_{13} \eta^{7/2} & \mu_{23}\eta^{2}  &  \eta^{5/2}
\end{array} \right) b_2 M_R
\eeq
 with $M_R = \eps' M_U = \sqrt{\eta} M_U $.

Having defined the Dirac and heavy Majorana mass matrices for the neutrinos, it
is straightforward to obtain the light Majorana mass matrix
from the see-saw formula \beq \label{seesaw}
 m_\nu = - m_{D\nu} M_N^{-1} m_{D\nu}^T
\eeq at the GUT scale.


\subsection{Setting the expansion parameters}

Given the fermion mass textures in terms of the $U(1)_X$ charges and expansion
parameters, we need now to determine the values of the latter in order to obtain
consistency with the low energy experimentally known quantities (masses and
mixing angles).
Note that the coefficient $b_2$ defined in Eq.~(\ref{rel}) will determine the
overall neutrino mass scale through Eq.~(\ref{MRHN}).

Consistency with the measured values of quark masses and mixings fixes the value
of $\eta\approx 5\times 10^{-2}$: for example the CKM mixing angle $\theta_{12}$
is given by $\sqrt{\eta}\approx 0.22$ up to small corrections~\cite{Dent:2004dn}.
Hence the ratio of the SU$(4)$ breaking scale $M_G$ to the fundamental scale
$M_U$ is also fixed through $\frac{M_G^2}{M_U^2} =\sqrt{\eta} \approx 0.22$: the
Pati-Salam group is unbroken over only a small range of energy. We perform a
renormalization group analysis in order to check the consistency of this
prediction with the low-energy values of the gauge couplings $\alpha_s$,
$\alpha_{em}$ and the weak mixing angle $\sin^2\theta_W$ \cite{Eidelman}
 \beq
 \sin^2\theta_W =0.23120,\ \alpha_3 = 0.118\pm 0.003,\ a_{em} = \frac{1}{127.906}.
 \nonumber
 \eeq
 At this point it should be noted that in our approach we assume that the
 underline theory is characterized by a single coupling constant.
 In D-brane constructions however, the gauge
couplings do not necessarily satisfy the usual unification condition. The reason
is that in this case, the volume of the internal space enters between gauge and
string couplings, thus the actual values of the gauge couplings may differ at the
unification scale. Nevertheless, it is possible that some internal volume
relations allow for a partial unification~\cite{Leontaris:2000hh}.
 If the underlying model at $M_U$ has a single unified gauge
 coupling, then $M_G$ is fixed to be just below the unification
 scale according to the analysis of gauge coupling unification
 in the MSSM. Because of this fact, the low energy measured range
 for $\alpha_3$ affects the unification of the gauge couplings.
 Thus, we add the following extra states
 \beq
 h_L = (1,2,1),\ \  h_R=(1,1,2)
 \nonumber
 \eeq
 which are usually present in a string spectrum~\cite{Antoniadis:1988cm}.
 It turns out that we need 4 of each of these extra states for $M_G = 9.32
 \times 10^{15}$ GeV to be consistent with the value of $\eps'$ deduced
 from quark mass matrices.

 \begin{figure}[!h]
\begin{minipage}{18pc}
\hbox{\hspace{-0.25in}
\includegraphics[scale=.83]{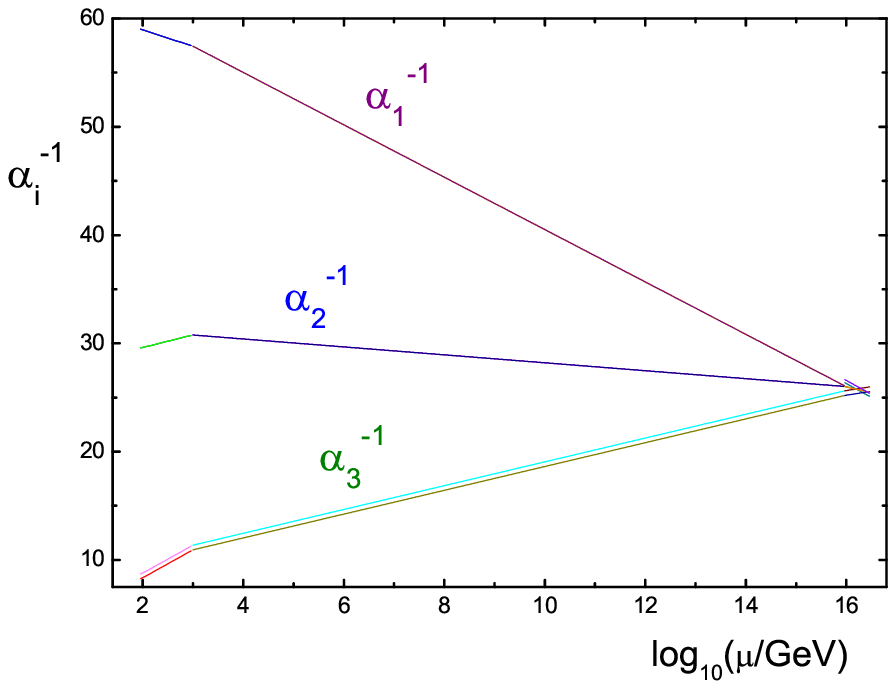}
 }
 \caption{\label{ggevo} Evolution of the gauge couplings. The two lines for
$\alpha_3$ indicate the range of initial conditions at $M_Z$.}
\end{minipage}
\hspace{2pc}%
\rsb{1.0ex}{\begin{minipage}{18pc}
 \hbox{\hspace{-0.4in}
\includegraphics[scale=.82]{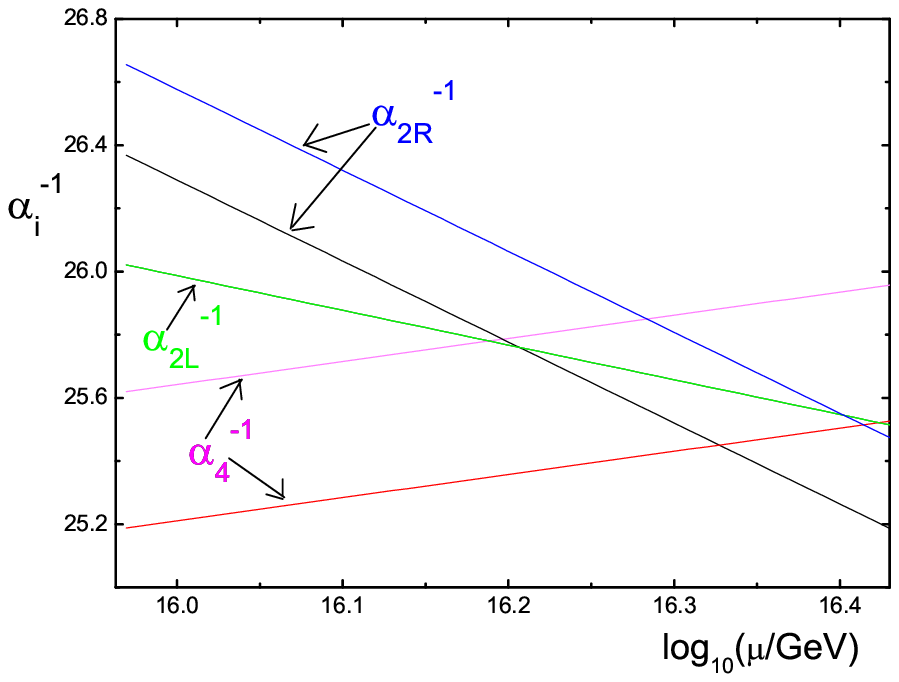}
 }
\caption{\label{closeup} Close-up of the gauge couplings in the
  Pati-Salam energy region.}
\end{minipage}}
\end{figure}
\begin{figure}[!h]
\includegraphics[scale=.88]{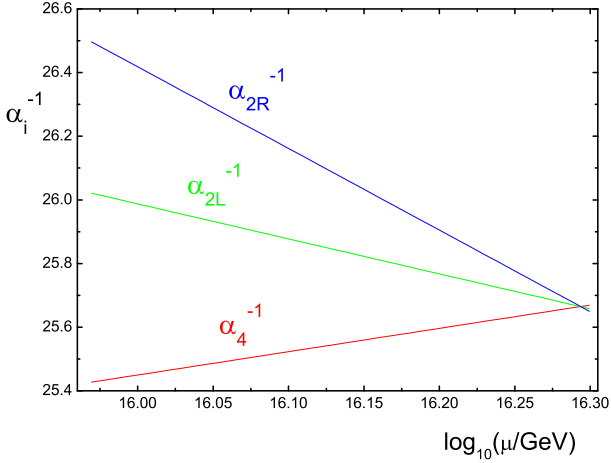}
 \rsb{4.5ex}{\begin{minipage}[b]{16pc} \caption{\label{psuni} The unification point of Pati-Salam gauge couplings for
 $\alpha_3 (M_Z)= 0.1176$.}
\end{minipage}}
\end{figure}

 In Figure \ref{ggevo}
 we plot the evolution of the gauge couplings from $M_Z$ to $M_U$.
  In Figure \ref{closeup}
 we show in more detail the evolution of the gauge couplings in the
 Pati-Salam energy region. The two bands for the $\alpha_4$
 and $\alpha_{2R}$ couplings are due to strong coupling
 uncertainty  at $M_Z$. For $\alpha_3 (M_Z)=0.1176$, as can
 be seen from Figure~\ref{psuni}, we obtain $M_U=1.96 \times 10^{16}$ GeV.

The remaining parameters to be determined are $b_1$, $b_2$ and $\mu_{ij}$ of the
right handed neutrino mass matrix.  We find that $M_N$ is proportional to $b_2$,
$M_N\approx b_2 M_R M_N' $, thus $b_2$ is related to the scale of the light
matrix $m_{\nu}$. Also,
 the choice $b_1 \approx 1.1$ leads to agreement with the data, while
implying $\eps = 1.1 \sqrt{\eta} \approx 0.25$.

\section{Running of neutrino masses and mixing angles}

The comparison between the high energy theoretical prediction and the
experimental observation is performed using the renormalization group equations.
The mass matrix for the light neutrinos is an outcome of the see-saw mechanism
and the effects induced by the RGEs are very important. Indeed, the low energy
neutrino data could be considerably different from the results at the see-saw
scale due to RG corrections above and below the see-saw thresholds.
 The
 running of neutrino
 masses and mixing angles has been extensively discussed for
 energies below the seesaw scales~\cite{below,Antusch:2002rr,Antusch:2003kp}
 as well as above~\cite{above,Antusch:2005gp,Mei:2005qp}. The
 running of the effective neutrino mass matrix $m_{\nu}$ above and
 between the see-saw scales is split into two terms,
 \beq
    m_\nu \,=\,
    -\frac{v^2}{4} \left(
    \, \accentset{(n)}{\kappa} +
    2 \, \accentset{(n)}{Y}_\nu \accentset{(n)}{M}^{-1}
     \accentset{(n)}{Y}_\nu^T \right) \;.
 \label{mnr}
 \eeq
where $\kappa$ is related to the coefficient of the effective 5 dimensional
operator $LLh_uh_u$, $(n)$ labels the effective field theories with $M_n$ right
handed neutrino integrated out ($M_n \ge M_{n-1} \ge M_{n-2}, \ldots$)
 and $ \accentset{(n)}{Y}_\nu$ are the neutrino couplings
 at an energy scale $M$ between two RH neutrino masses $M_n \ge M
 \ge M_{n-1}$, while $\accentset{(n)}{Y}_\nu=0$ below the
 lightest RH neutrino mass. These effective parameters
 govern the evolution below the highest seesaw scale
 and obey the differential
 equation~\cite{below,Antusch:2002rr,Antusch:2003kp}
 \beq
 \label{eq} 16 \pi^{2} \frac{d \effn{n} X}{dt} = ( Y_e Y_e^\dagger + \effn{n}
Y_\nu \effn{n} Y_\nu^\dagger)^T  \effn{n} X + \effn{n} X ( Y_e Y_e^\dagger  +
\effn{n} Y_\nu \effn{n} Y_\nu^\dagger)^T
 \\
 + (2 \mrm{Tr} (\effn{n} Y_\nu \effn{n} Y_\nu^\dagger +
3 Y_u Y_u^\dagger) -6/5 g_1^2 -6 g_2^2) \effn{n} X
\eeq
 where $X=\kappa$, $Y_\nu M^{-1} Y_\nu^T$.
 The RGEs have been solved both numerically and also
 analytically~\cite{Antusch:2003kp,Antusch:2005gp,Mei:2005qp}.
 Numerically, below the lightest heavy RH neutrino mass large
 renormalization effects can be experienced only in the case of
 degenerate light neutrino masses for very large $\tan\beta$~\cite{Ellis:1999my,Antusch:2003kp}.
 Above this mass things are more complicated due to the
 non-trivial dependence of the heavy neutrino mass couplings,
 unless $M_{\nu^c}$ is diagonal. For the leptonic mixing
 angles, in the case of normal hierarchy relevant to our
 model, one expects negligible effects for the solar mixing
 angle while $\theta_{13}$ and $\theta_{23}$ are expected
 to run faster~\cite{Ellis:2005dr}.

 On the other hand, studying the analytical expressions
 obtained after approximation, exactly the opposite behavior
 is predicted and the solar mixing angle receives larger
 renormalization effects than $\theta_{13}$ or $\theta_{23}$.
 However, possible cancellations may occur and enhancement or suppresion factors
 may appear, thus the numerical solutions may differ considerably
 from these estimates.

In our string-inspired model the Dirac and heavy Majorana mass matrices at the
unification scale are parametrized in terms of order-1 superpotential
coefficients $\mu_{ij}(M_U)$
whose exact numerical values are not known. The flavour structure at the
unification scale might also be different from that at the electroweak scale
$M_Z$. Thus, even if the Yukawa parameters are determined at $M_Z$, to understand
the structure of the theory at $M_U$, and consequently any possible family
symmetry, we would certainly need the parameter values at $M_U$.

\begin{table}
\caption{\label{nummuij} Numerical values of parameters $\mu_{11}$, $\mu_{12}$,
$\mu_{13}$, $\mu_{22}$, $\mu_{23}$ at $M_G$.}
\begin{center}
\begin{tabular}{cccccc}
\br Solution & $\mu_{11}$& $\mu_{12}$ & $\mu_{13}$ & $\mu_{22}$ & $\mu_{23}$ \\
\mr
1.  & 0.10535 & 0.10972 & 0.86012 & 0.10491 & 0.91014 \\
2.  & 0.11939 & 0.10954 & 0.80912 & 0.10683 & 0.93832 \\
3.  & 0.10392 & 0.11787 & 0.97796 & 0.10512 & 0.98749 \\
4.  & 0.09143 & 0.10962 & 0.87616 & 0.10063 & 0.93798 \\
5.  & 0.12697 & 0.12745 & 0.99860 & 0.11652 & 0.99980 \\
6.  & 0.10920 & 0.09638 & 1.00975 & 0.10238 & 0.93561 \\
7.  & 0.10124 & 0.11682 & 0.98568 & 0.10688 & 0.99156 \\
8.  & 0.12358 & 0.09514 & 0.99580 & 0.10434 & 0.95646 \\
9.  & 0.13006 & 0.11973 & 1.02235 & 0.10378 & 0.89460 \\
10. & 0.12665 & 0.12137 & 1.00029 & 0.10695 & 0.91578 \\
\br
\end{tabular}
\end{center}
\end{table}

In this section we study the renormalization group flow of the neutrino mass
matrices in a ``top-down'' approach from the Pati-Salam scale $M_G$ to the weak
scale. We choose sets of values of the undetermined order 1 coefficients at the
high scale and run the renormalization group equations down to $M_Z$ where we
calculate $\Delta m_{\nu_{ij}}^{2}$ and $\theta_{\nu_{ij}}$ and compare them with
the experimental values. Study of a bottom-up approach has been
performed~\cite{Ellis:2005dr} and we will compare our results to this work. The
renormalization group analyses of the neutrino parameters, successively
integrating out the right handed neutrinos, is performed using the software
packages REAP/MPT~\cite{Antusch:2005gp}.

We generate appropriate numerical values for the coefficients $\mu_{11}$,
$\mu_{12}$, $\mu_{13}$, $\mu_{22}$, $\mu_{23}$, so that after the evolution of
$m_{\nu}$ to low energy we obtain values in agreement with the experimental data.
The coefficient $\mu_{33}$ is set to unity (which can always be done by adjusting
the value of $b_2$). Experimentally acceptable solutions can be seen in Table
\ref{nummuij}. In Table \ref{dmthetanum} we present the resulting values of
$\theta_{ij}$ and $\Delta m_{ij}^2$ at the scale $M_Z$. The mass-squared
differences lie in the ranges
 $\Delta m_{atm}^2=[1.33-3.39]\times 10^{-3}$eV$^2$, $\Delta
m_{sol}^2=[7.24-8.85]\times 10^{-5}$eV$^2$. These are consistent with the
experimental data ${\Delta m_{atm,exp}^2}= [1.3-3.4]\times 10^{-3}$eV$^2$ and
${\Delta m_{sol,exp}^2}= [7.1-8.9]\times 10^{-5}$eV$^2$. The mixing angles are
also found in the allowed ranges $\theta_{12}=[29.4-37.6]$,
$\theta_{23}=[36.9-51.0]$ and $\theta_{13}=[0.86-12.50]$.

\begin{table}
 \caption{\label{dmthetanum} Values of the physical parameters $\Delta
m^2_{atm}$, $\Delta m^2_{sol}$, $\theta_{12}$, $\theta_{13}$, $\theta_{23}$ at
$M_Z$ (mass units eV$^2$).}
\begin{center}
\begin{tabular}{cccccc}
 \br Solution & $\Delta m_{atm}^2 (M_Z) \cdot 10^{3}$ & $\Delta m_{sol}^2 (M_Z) \cdot 10^{5}$ &
$\theta_{12} (M_Z)$ & $\theta_{13} (M_Z)$ & $\theta_{23} (M_Z)$ \\
\mr
1.  & 2.7149 & 7.9621 & 29.442 & 3.9859 & 44.114 \\
2.  & 2.3145 & 7.9514 & 34.289 & 12.507 & 51.047 \\
3.  & 1.8978 & 8.6141 & 30.560 & 0.8656 & 46.230 \\
4.  & 3.0062 & 8.3217 & 34.347 & 1.8512 & 44.333 \\
5.  & 3.3905 & 7.2468 & 30.245 & 2.9355 & 36.900 \\
6.  & 3.2459 & 7.5351 & 34.296 & 1.3701 & 46.947 \\
7.  & 2.0171 & 7.9464 & 34.432 & 1.0086 & 50.279 \\
8.  & 1.3321 & 7.9060 & 37.646 & 6.1490 & 43.067 \\
9.  & 2.4867 & 8.8561 & 29.592 & 5.6007 & 42.970 \\
10. & 2.1652 & 7.8869 & 29.189 & 3.1512 & 37.220 \\
\br
\end{tabular}
\end{center}
\end{table}

\begin{figure}[!h]
\begin{minipage}{18pc}
\hbox{\hspace{-0.25in}
\includegraphics[scale=0.82]{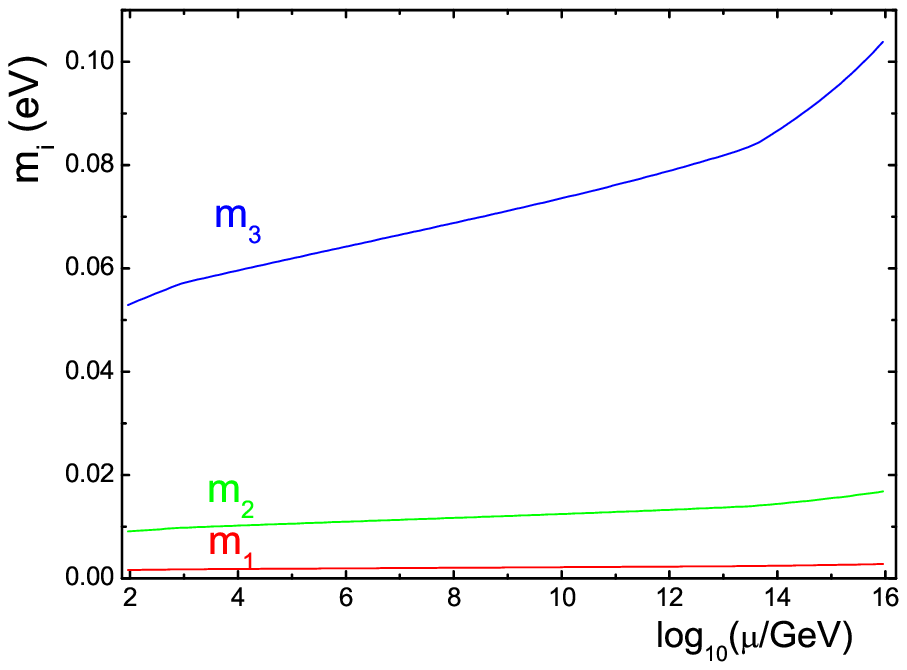}
 }
\caption{\label{lightneum} The running of the light neutrino masses.}
\end{minipage}\hspace{2pc}%
\rsb{-0.35ex}{\begin{minipage}{18pc}
 \hbox{\hspace{-0.5in}
\includegraphics[scale=0.82]{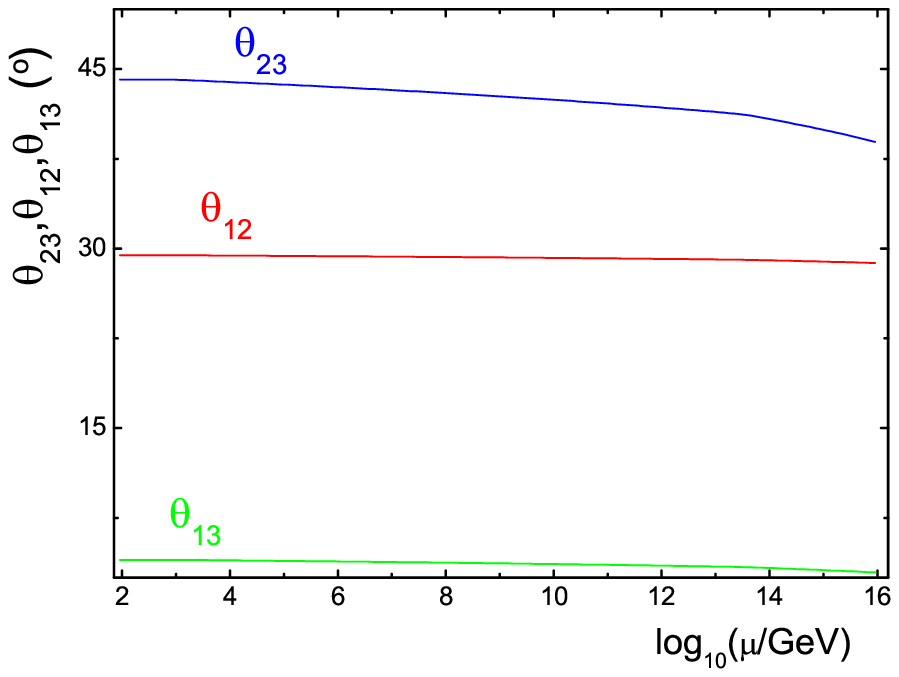}
 }
\caption{\label{anglesevolatex} The evolution of the mixing angles.}
\end{minipage}}
\end{figure}

In figure \ref{lightneum} we plot the running of the three light neutrino
Majorana masses ($m_1<m_2<m_3$) in the energy range $M_G-M_Z$. The initial (GUT)
neutrino eigenmasses are all larger than their low energy values. Significant
running is observed mainly for the heaviest eigenmass $m_{\nu_3}$. For
experimentally acceptable mass-squared differences $\Delta m_{\nu_{ij}}^2$ at
$M_Z$, in all cases their corresponding values at the GUT scale lie out of the
acceptable range. In this scenario with hierarchical light neutrino masses, we
find that large renormalization effects occur above the heavy neutrino threshold
since the Yukawa couplings $Y_{\nu}$ are large and the second term in (\ref{mnr})
dominates. Also, since $m_{\nu_1}< \sqrt{\Delta m^2_{sol}}$, the solar angle
turns out to be more stable compared to the running of the $\theta_{23}$, as can
be seen in Figure \ref{anglesevolatex}. These results are in agreement with the
findings of~\cite{Ellis:2005dr}.

%
\begin{table}
\caption{\label{dmthetanumatmgut} Values of the physical parameters $\Delta
m^2_{atm}$ and
 $\Delta m^2_{sol}$ at $M_G$; the effective neutrino mass $\vev{m}$
 related to $\beta\beta_{0 \nu}$ decay; and the parameter $b_2$
 which determines the scale of the light matrix $m_{\nu}$.}

\begin{center}
\begin{tabular}{ccccc} \br Solution & $\Delta m_{atm}^2(M_G) \cdot 10^{3}$ &
$\Delta m_{sol}^2(M_G)\cdot 10^{5}$ & $\langle m \rangle$ & $b_2$\\
 \mr 1.  & 10.5294 & 27.6096 & 0.00361 &3.41\\
  2.  & 8.1987 & 30.1333  & 0.00633 &2.88\\
  3.  & 7.2533 & 31.6108 & 0.00607 &1.50\\
  4.  & 11.7749 & 30.4352 & 0.00753 &1.28\\
  5.  & 14.093 & 23.7416 & 0.00346 &2.85\\
  6.  & 12.347 & 28.2642 & 0.00822 &1.26\\
  7.  & 7.4256 & 30.855 & 0.00821 &1.23\\
  8.  & 5.2910 & 29.8775 & 0.00852 &1.17\\
  9.  & 9.74656 & 30.5418 & 0.00379 &3.64\\
  10. & 8.99695 & 25.8838 & 0.00327 &3.42\\
\br
 \end{tabular}
\end{center}
\end{table}

\begin{figure}[!h]
\rsb{1.5ex}{\begin{minipage}{18pc}
 \hbox{\hspace{-0.25in}
\includegraphics[scale=0.8]{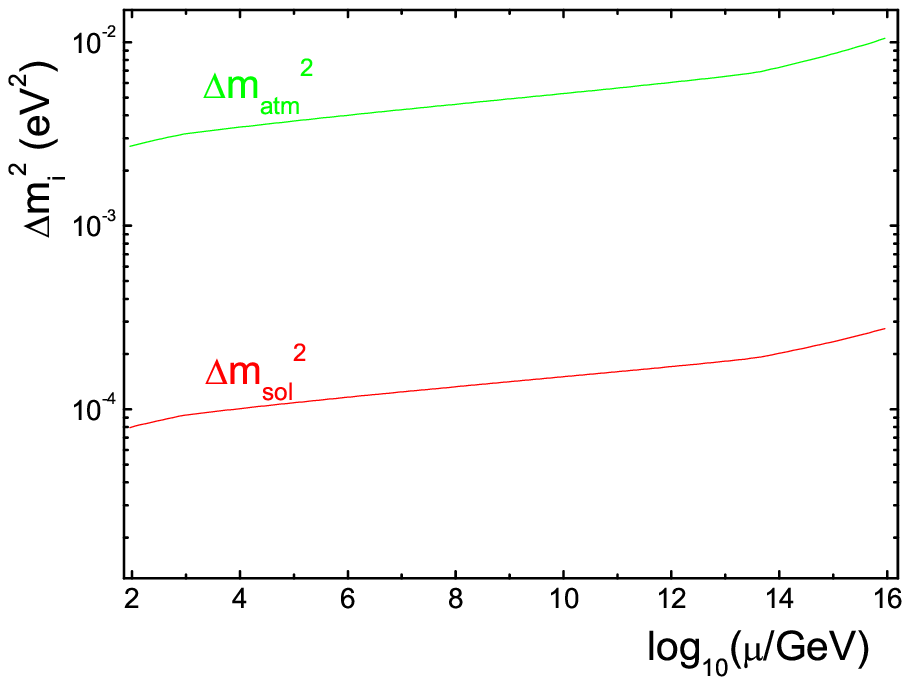}
 }
\caption{\label{splittingsevolatex} Running of $\Delta m^2_{sol}$ and $\Delta
m^2_{atm}$.}
\end{minipage}}\hspace{3.5pc}%
\begin{minipage}{18pc}
 \hbox{\hspace{-0.75in}
\includegraphics[scale=0.8]{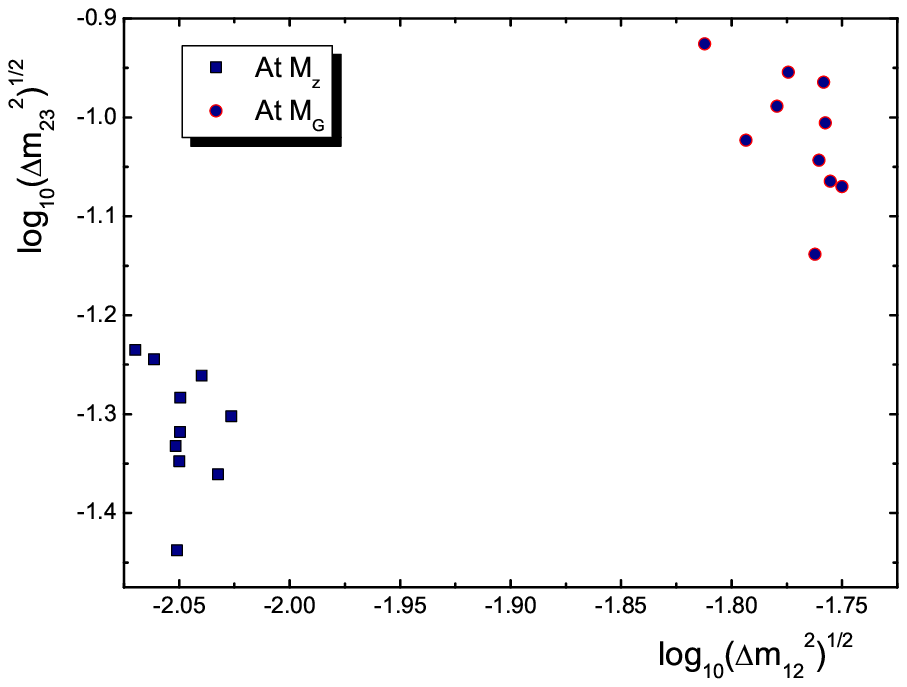}
 }
 \caption{\label{dmmgmz} $\Delta m^2_{12}$ and $\Delta
m^2_{23}$ at $M_Z$ \\ and at $M_G$. }
\end{minipage}
\end{figure}

The evolution of the atmospheric and solar mass differences is been depicted in
figure \ref{splittingsevolatex}. In figure \ref{dmmgmz} we plot the distribution
$\Delta m_{atm}^2$ versus $\Delta m_{sol}^2$ at the two scales $M_G$ (Table
\ref{dmthetanumatmgut}) and $M_Z$ for the ten models of Table \ref{nummuij}. We
find that the hierarchy of the neutrino masses at the Pati-Salam breaking scale
tends to be greater than that at low energies. Several models predict $\Delta
m_{23}^2/\Delta m_{21}^2$ out of the experimental range at $M_G$, although after
the running at $M_Z$ they are consistent with the data.

Finally, we check the predictions of our model for the effective neutrino mass
parameter relevant for $\beta\beta_{0 \nu}$ decay. This parameter can be written
in terms of the observable quantities as
 \beq
 |\langle m \rangle| = \left|(m_1 \cos^2\theta_\odot  + e^{i \alpha}
\sqrt{\Delta m_{sol}} \sin^2 \theta_\odot)
\cos^2\theta_{13}\right. \\
+ \left. \sqrt{\Delta m_{atm}} \sin^2\theta_{13}
 e^{i\beta} \right|.
\label{mef}
 \eeq

In the last column of Table (\ref{dmthetanumatmgut}) the $\beta\beta_{0
\nu}$-decay predictions are presented for solutions 1-10. Many current
experiments attempt to measure this quantity~\cite{Petcov:2005yq}; the best
current limit on the effective mass is given by the Heidelberg--Moscow
collaboration~\cite{heidmo} \beq \label{eq:current} \langle m \rangle \le 0.35\,
z~ \mbox {eV}, \eeq where the parameter $z={\cal O}(1)$ allows for uncertainty
arising from nuclear matrix elements.

In a recent analysis of neutrinoless double beta decay~\cite{Choubey:2005rq} the
allowable range of the effective mass parameter was given for specific scenarios.
In the case of the normal hierarchy the bounds are \beq 0 < \langle m \rangle <
0.007\,\mbox{eV} \eeq thus our results are in the experimentally acceptable
region.

%


\section{Conclusions}

In this talk, we studied the running of neutrino masses and mixing angles in a
supersymmetric string-inspired SU$(4)\times$SU$(2)_L\times
$SU$(2)_R\times$U$(1)_X$ model. An accurate description of the low energy
neutrino data forced us to introduce two singlets charged under the $U(1)_X$,
leading to two expansion parameters. The mass matrices are then constructed in
terms of three expansion parameters \beq \eps \equiv \frac{\vev{\phi}}{M_U},\ \
\eps' \equiv \frac{\vev{H\bar{H}}}{M_U^2},\ \ \eps'' \equiv
\frac{\vev{\chi}}{M_U}, \eeq where $\phi$ and $\chi$ are singlets and $H$,
$\bar{H}$ the SU$(4)\times$SU$(2)_R$-breaking Higgses.
The model is simplified by
the fractional U$(1)_X$ charges of $H$ and $\chi$, which ensure that the
parameter $\eps''$ only appears as a prefactor to the heavy Majorana neutrino
masses.

The expansion parameter arising from the Higgs v.e.v.'s defines the ratio of the
$SU(4)$ breaking scale $M_G$ to the unification scale $M_U$: we performed a
renormalization group analysis of gauge couplings under this constraint and found
successful unification with the addition of extra states usually present in a
string spectrum.

Assuming that only the third generation of quarks and charged leptons acquire
masses at tree level and under a specific choice of $U(1)_X$ charges as well as
Clebsch factors, we examined the implications for the light neutrino masses
resulting from the see-saw formula. We found that the light neutrino mass
spectrum is hierarchical and that the mass hierarchy tends to be larger at the
GUT scale than at $M_Z$ due the renormalization group running. The solar mixing
angle $\theta_{12}$ is stable under RG evolution while larger renormalization
effects are found for the atmospheric mixing angle $\theta_{23}$ and
$\theta_{13}$, always with their values at $M_Z$ in agreement with experiment.

\ack \

This research was funded by the program `HERAKLEITOS' of the Operational Program
for Education and Initial Vocational Training of the Hellenic Ministry of
Education under the 3rd Community Support Framework and the European Social Fund.

\section*{References} \

\end{document}